\documentclass[journal,twocolumn,10pt]{IEEEtran}
\pdfoutput=1
\usepackage{cite}
\usepackage{amsmath}
\usepackage{color}
\usepackage{subcaption}
\usepackage{multicol}

\ifCLASSINFOpdf
    \usepackage[pdftex]{graphicx}
    \DeclareGraphicsExtensions{.pdf,.jpeg,.png}
\else
    \usepackage[dvips]{graphicx}
    \DeclareGraphicsExtensions{.eps}
\fi

\usepackage{geometry}
\include{pics}

\begin{document}
\title{Multipath Communication with Finite Sliding Window Network Coding for Ultra-Reliability and Low Latency
}
\author{
	\IEEEauthorblockN{
		Frank Gabriel, 
		Anil Kumar Chorppath,
		Ievgenii Tsokalo, and
		Frank H.P. Fitzek
	}\\
	\IEEEauthorblockA{
        Deutsche Telekom Chair of Communication Networks, TU Dresden, 01062 Dresden, Germany\\
        $[$frank.gabriel$|$anil.chorppath$|$ievgenii.tsokalo$|$frank.fitzek$]$@tu-dresden.de
    }
}

\maketitle
\begin{abstract}
We use random linear network coding (RLNC) based scheme for multipath communication in the presence of lossy links with different delay characteristics to obtain ultra-reliability and low latency. A sliding window version of RLNC is proposed where the coded packets are generated using packets in a window size and are inserted among systematic packets in different paths. The packets are scheduled in the paths in a round robin fashion proportional to the data rates.  We use finite encoding and decoding window size and do not rely on feedback for closing the sliding window, unlike the previous work. Our implementation of two paths with LTE and WiFi characteristics shows that the proposed sliding window scheme achieves better latency compared to the block RLNC code. It is also shown that the proposed scheme achieves low latency communication through multiple paths compared to the individual paths for bursty traffic by translating the throughput on both the paths into latency gain.
\end{abstract}

\section{Introduction}

Multipath transmission gives high promise for the 5th generation (5G) networks to solve the adaptive requirements of data rate, latency, and reliability. Multipath comes naturally in the ecosystem where there is a convergence of wired and wireless (licensed, unlicensed and device-to-device) interfaces to a single device. Many emerging use cases such as V2X communication and Tactile Internet puts stringent requirements on latency and reliability. In \cite{Multipath} multipath communication through multiple communication interfaces has been proposed as an important enabler for ultra-reliable low latency communication (URLLLC). We consider in this paper streaming applications where there is a high requirement on latency and evaluates an LTE-WiFi integration solution. 

\begin{figure}[!t]
	\centering
	\includegraphics[width=0.95\columnwidth]{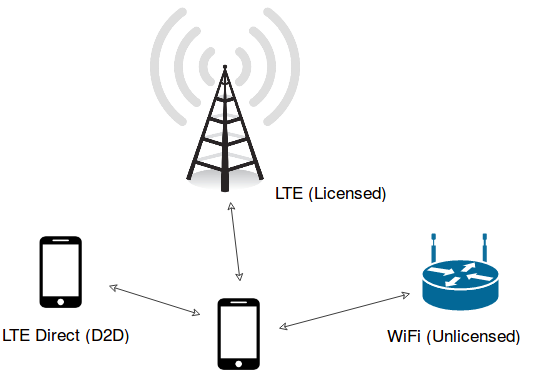} 
	\caption{Device with multiple interfaces}
	\label{fig:paths}
\end{figure}

In case of LTE-WiFi aggregation solutions led by 3rd Generation Partnership Project (3GPP), the aggregation takes place at radio link layer. However, for the Proxy-based LTE-WiFi aggregation solutions led by Internet Engineering Task Force (IETF) this takes place at transport layer with Multipath TCP (MPTCP). The purpose of MPTCP is to transmit data using as many paths as possible while still working with the existent Internet environment and extended the legacy TCP to establish multiple paths to send data concurrently through them. Motivated with this direction, we would like to propose multipath transmission with sliding window network coding for low-latency and reliable communication. Figure \ref{fig:paths} shows a network with three paths, LTE , WiFi and LTE Direct (a Device-to-Device (D2D) path) to a single device.
 

Instead of duplicating packets in multiple paths for high reliability, we propose to use a sliding window network code for forward error correction (FEC). Sliding window network coding is a low delay version of random linear network coding (RLNC). The trade-off between reliability, delay and throughput can be adjusted with the code rate and the encoding window size.


Scheduling by delaying packets based on the delay difference in paths is not feasible due to the dynamical nature of wireless links. We would like to tackle the effect of delay asymmetry in different paths in multipath transmission. For multipath communication compared to the single path, the coding window size needs to be higher to compensate for the delay asymmetry.

The following are the contributions of the paper. It shows the usefulness of transmission on multiple paths (with different characteristics) compared to using only fastest path for reducing latency in the presence of source with bursty traffic. The ineffectiveness of block codes to compensate for the delay asymmetry is also shown. We obtain the decoding window size of the sliding window RLNC based on the delay asymmetry, to include the packets from the slower path in decoding, and the encoding window size.  The effect of code rate on the performance of block codes and sliding window is shown and the sliding window is shown to perform better.

The paper is organized as follows. In Section \ref{sec:soa} we describe the current state of the art for multipath communication and the difference to our work. This is followed by a basic description of sliding window network coding and its multipath extension in Section \ref{sec:setup}. The section also analyzes the code rate, window size, delay and packet loss trade-off. Section \ref{sec:Eval} gives the simulation results and the paper is concluded in Section \ref{sec:Conc}.

\section{State-of-the-art }
\label{sec:soa}

Multipath communication using cellular and D2D links are proposed for 5G in \cite{MultipathD2D}.
In \cite{Multipath}, packet duplication in three interfaces, LTE, HSPA and WiFi, is shown to perform better compared to individual paths for URLLC communication. Interleaving network coded packets with systematic packets is much more efficient than packet duplication. The delay-throughput trade-off is studied in \cite{Delay-Throughput} where channel coding is done on physical layer and multipath routing is done on higher layer. In contrast, our proposed RLNC scheme does not require modifications of the physical layer and can be deployed on current hardware.

An analysis of decoding performance of RLNC for broadcast is given in \cite{DecodingDelayPerformance}. Network coded multipath communication has been studied in \cite{5502204,6193502,4745146}, focusing on the achieved throughput. Sliding window RLNC was proposed in \cite{KarzandMurielAllerton} as a low delay modification to block RLNC codes which assumes an infinite encoding window size or a sliding
window of dynamic size, which is closed only by feedback. A finite sliding window modification without feedback was evaluated with bursty traffic in \cite{NoGeneration,crlnc}. A similar study\cite{RocaSW} shows that a sliding window based convolution code is better compared to Reed-Solomon codes for low latency and used decoding window size greater than the encoding window size for higher reliability.

A multipath extension of \cite{KarzandMurielAllerton} is proposed in \cite{CloudM16} and it is shown that the increase in overall in-order delay due to communication over multiple paths is insignificant when considering the potential throughput gains. Again, feedback is used to close the sliding window. We propose a scheme without feedback and evaluate it in a low latency multimedia streaming application. In addition, we analyze the effect of delay asymmetry on the decoding coding window size. We also carry out a comparison between sliding window code and block codes for the latency performance.

A similar algorithm called  Stochastic Earliest Delivery Path First(S-EDPF) a generalization of EDPF which takes into account uncertainty and time-variation in path delays is proposed in \cite{Garcia-SaavedraTMC} where the coded packets are sent only on one path. 

A new multipath scheduler aiming at reducing the data chunk download time in the context of MPTCP for mobile communication using WiFi/LTE is given in \cite{Guo:Multipath:Subflow}. A traffic aware scheduling strategy in wireless networks for bursty traffic is given in \cite{Burstytraffic} which proposed the need for network control for bursts in traffic.
 
Values for delay and packet loss of WiFi and LTE connections is provided in the data trace in \cite{Partov:LeithTrace}. Another data trace on WiFi traffic is given in \cite{deng2014wifi}. We use the traces as basis for the channel model in our evaluation.
 

\section{Sliding Window RLNC for Multipath Communication}
\label{sec:setup}

RLNC linearly combines source (original data) symbols over a Galois field $GF (q)$ to create coded symbols. Source symbols can be represented as row vectors of elements of $GF (q)$. In a systematic code, the source symbols are sent with additional coded symbols as redundancy. Block-based RLNC groups source symbols into the equally sized block of $g$ subsequent source symbols each. Such a block is also called a generation, and $g$ is called the generation size. Block-based RLNC creates typically multiple coded symbols in a batch. The coding step can be expressed as matrix multiplication of a coefficient matrix $C$ (consisting of $g + n_c$ rows and g columns of
coefficients in $GF (2^8 )$) with the source symbol matrix $X$.

\begin{figure*}[!t]
	\centering
	\includegraphics[width=0.9\textwidth]{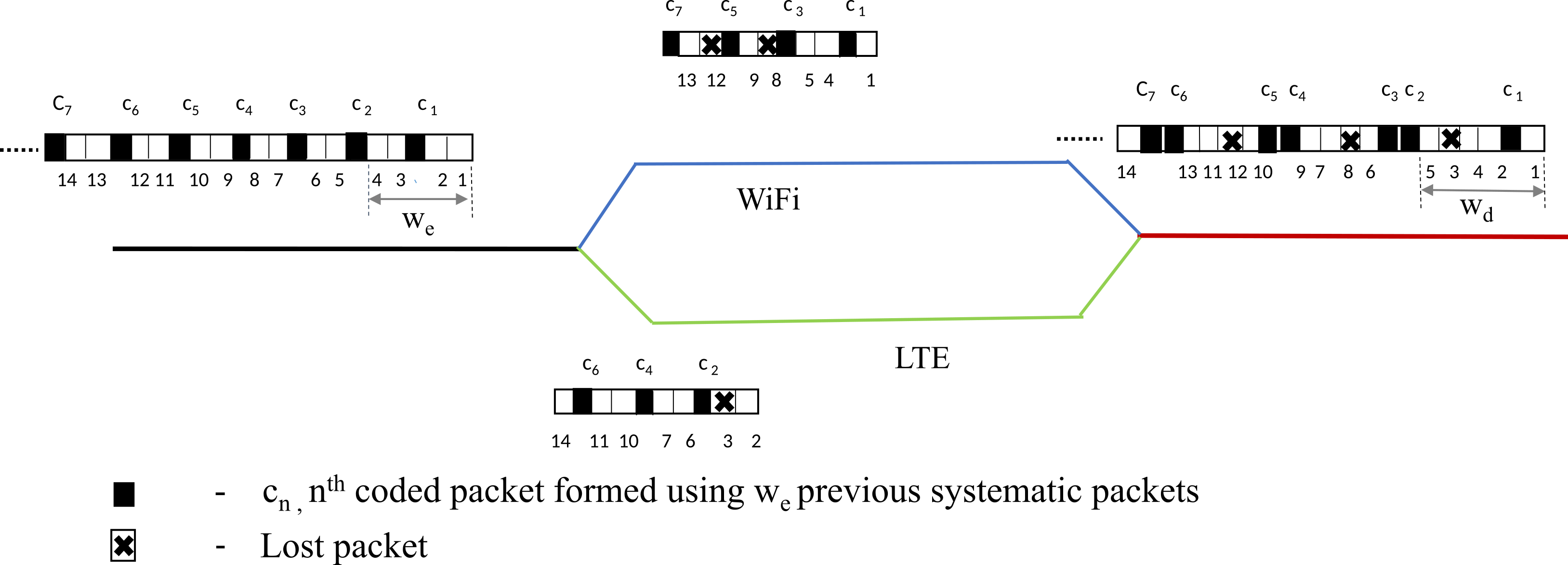} 
	\caption{Multipath  sliding window scheme with two paths; LTE and WiFi. LTE path has 1 packet delay more than WiFi path. LTE path has half the loss probability compared to WiFi path.}
	\label{fig:paths1}
\end{figure*}

Sliding window RLNC does not group the source symbols into artificial blocks of generations.
Instead, all source symbols within a consecutive sequence are considered for creating coded symbols.
This consecutive sequence of source symbols is referred to as the sliding window. 

Appending new source symbols to the bottom (end) of this sequence of
symbols is commonly referred to as opening the window. In
contrast, removing source symbols from the top (beginning) of the sequence is referred to as closing the window. In practice, the window is usually opened when new source symbols arrive at the encoder. Closing the window can be done with feedback \cite{KarzandMurielAllerton} or automatically \cite{crlnc}.
The code rate is commonly controlled by the number of source symbols $n_s$ that are sent before one coded symbol,
i.e., the code rate is $R = \frac{n_s}{(n s + 1)}$. More specifically, for systematic sliding window RLNC, $n_s$ uncoded source
symbols are sent followed by one coded symbol. Therefore, $g$ is replaced by $n_s$ and $n_c=1$ for sliding window RLNC. 

For finite sliding window RLNC, we use a finite sliding encoding window of size $w_e , w_e \geq n_s$. The finite sliding window encoder
uses the last $w_e$ source symbols to create a coded symbol, while the code rate is fixed at a prescribed $R = \frac{n_s}{(n s + 1)}$
as in conventional sliding window RLNC. \figurename~\ref{fig:paths1} gives an overview of the proposed scheme.

\subsection{Code rate Selection}
\label{sec:CR}

Code rate at the sender is same for generating packets for both paths. The code rate needs to be modified for sending coded packets on all the paths for compensating losses on all the paths \cite{CloudM16}. 
$$ \sum_i (1-\epsilon_i) (1-\epsilon_i- R_i)r_i > 0, $$
where $\epsilon_i$ is the loss ratio in path $i$.
The code rate for two paths with same data rate is given by
$$(1-\epsilon_1) (1-\epsilon_1- R_1) + (1-\epsilon_2) (1-\epsilon_2- R_2) > 0 $$
For same code rate in both paths,
\begin{equation} \label{CRate}
R< 1- \frac{ \epsilon_1(1-\epsilon_1)+ \epsilon_2 (1-\epsilon_2) }{1-\epsilon_1+1-\epsilon_2} 
\end{equation}
 The equation (\ref{CRate}) provides the limit on code rate to ensure successful decoding in the presence of losses on two paths. Usually the code rate is $R< 1- \epsilon $ for one path.

For example, for $\epsilon_1= 0.1$ and $\epsilon_2= 0.05$, with coded paths in one path $R< 1-0.095(=0.905)$ (as in \cite{CloudM16}) and in two paths $R< 1-0.072(=0.928)$. We observe that there is more room for high code rate if coded packets are split between 2 paths.
With only single path available, the limit on code rate is $c< 0.90$ in the bad path (path 1) and $R< 0.95$ in the good path (path 1). For $\epsilon_1= 0.2$ and $\epsilon_2= 0.1$, with coded packets only on path 1, $R< 1-0.095(=0.732)$ (as in \cite{CloudM16}) and with coded packets in both the paths $R< 1-0.072(=0.853)$.

So far, the effect of finite window size for compensating the difference in delay on the paths is not considered. For the finite coding window, the possible code rates will be lower than the one obtained above. For different window sizes and coding rates, in the next section, we investigate the latency with two paths and the possible values of packet loss. For the finite sliding window, there will be some losses even within the bounds on code rate above, but the mean delay will be lower since the packets will not be queued up infinitely.

\subsection{Relation of delay, rate and loss probability}
\label{sec:buffer}

Let there are two parallel paths with delays $d_1$ and $d_2$ and $d_1 \leq d_2$.
The rate of transmission on the paths are $r_1$ and $r_2$.
With the code rate $R$ this network can transport a signal with rate $r_s$ if $r_s \leq R ( r_1 + r_2 )$.
The packets are sent in order as soon as a path is available.
As a special case, for $r_1 = r_2$ this results in round-robin scheduling.

\begin{figure}[!t]
	\centering
	\caption{Evaluation of the buffer size. In both cases, need buffering for data received over the path 1 for the duration $T=d_2 - d_1$}
	\label{fig:timings}
\end{figure}

Because of the delay difference $\Delta d = d_2 - d_1$ of the paths, the packets will not be received in the correct order.
The packets from the path with the lower delay need to be buffered.
The buffer size required to compensate for the delay is $B = r_1 \Delta d + 1$. 
As shown in \figurename~\ref{fig:timings}, the receiver buffer size $B$ in case with no coding can be evaluated as $B=r_1 \Delta d + 1$ independent on ratio between the transmission rates $r_1$ and $r_2$.

For RLNC, the packets cannot be immediately discarded when they are received in-order.
Instead, the packets could be necessary to decode future coded packets.
The decoder has to store all packets for a generation until the generation can be discarded.
The generation can only be discarded if all packets from the current and previous generations are decoded
or there are no more coded packets for the current and previous generations in flight.
From the second case, we derive the required buffer size in the decoder.

When a packet on the path with the high delay is received, the path with the low delay can already receive packets from a newer generation.
The difference in the generation number depends on the delay difference and the number of source packets sent on all paths.
If the less new information is sent the packets from the low-delay path are less ahead.
To be able to use all packets from the high-delay path the decoder has to store at least

\begin{equation}
B_G \geq \left\lfloor \frac{(r_1 + r_2) R \Delta d}{g} \right\rfloor + 1
\end{equation}

generations. As a result, the maximum number of packets in the decoding buffer is $g \cdot B_G$.

\begin{figure}[!t]
    \includegraphics{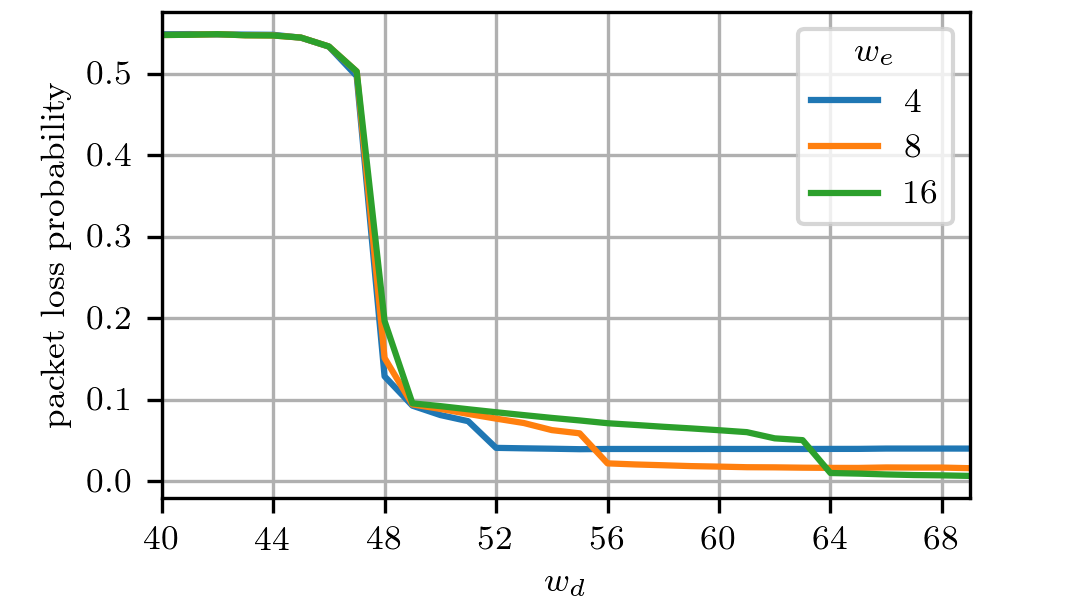}
    \caption{Simulation results for various encoding and decoding window sizes for a network with $r_1=r_2=3000$ packets per second, $e_1=e_2=0.1$, $c=0.8$, $r_s=R(r_1+r_2)$}
    \label{fig:window}
\end{figure}

For sliding window RLNC, the decoding buffer needs to at least contain all symbols to decode a received coded packet.
When a packet is received on the high-delay path, the most recent packet in the decoding window, received on the low-delay path, is $(r_1 + r_2) R \Delta d$ packets ahead.
Additionally, the decoding window requires at least $w_e$ packets to decode the packets from the high-delay path.
In total, the number of packets in the decoding window must be at least

\begin{equation}
    B \geq (r_1 + r_2) R \Delta d + w_e
\end{equation}

to use all packets from both paths.
\figurename~\ref{fig:window} demonstrates this based on simulation results.
Up until a decoding window size of 48, no packets from the high-delay path can be used and the loss is above 50\%.
At this point, the systematic packets from the high-delay path fit in the decoding window.
But only the systematic packets can be used and as a result, the packet loss probability is still around 10\%.
When the decoding window size is increased to 48 + $w_e$ also the coded packets can be used.
The packet loss probability is significantly reduced.
From this point forward increasing the decoding window size still reduces the packet loss probability, but only to a small extent.
This is also observed in \cite{RocaSW}.

\subsection{Proposed Multipath Scheme}

We propose a finite encoding and decoding window multipath sliding window scheme. For a low latency application, explicitly closing the window with feedback is not necessary and not useful. Closing the window can be inferred from the maximum acceptable delay and the round trip time for feedback is relatively high.

The network coded packets are generated in a systematic fashion. The sliding window and the code rate are selected for minimizing the delay and compensating for the errors as required by the application. To allow a fine-grained adjustment of the code rate, a credit based implementation is used. For each sent source packet the credit counter is incremented by $(R^-1-1)$. When the counter exceeds 1, a coded packet is generated and the counter is decremented.

The network coded packets are sent over both the paths in a round robin fashion and queued up at the receiver buffer for the decoding. Gaussian elimination is used for the decoding and the packets which cannot be decoded in the decoding window are dropped.

In the next section, we evaluate the performance of this scheme in the context of a real-time video stream.

\section{Evaluation}
\label{sec:Eval}
For a practical evaluation of sliding window in the multipath setting, we use a discrete event simulator.
The simulator creates an input stream of packets that are given to the encoder.
We designed the traffic generator based on the captured traffic from a web-camera that used MJPEG codec.
MJPEG compresses each frame individually.
This requires more bandwidth than more sophisticated video encoders but has lower encoding and decoding delay.
Because each frame is a compressed image, the best fitting traffic model is a deterministic sequence of packet bursts.
Each frame is compressed and sent out as a burst of packets.
The size of the frame, and thus the number of packets per burst, depends on the resolution and the quality of the compression.
For a 720p video stream, we observed bursts of 42 packets of size 1472.
The rate of the bursts is given by the frame rate of the video.
We used 60 frames per second.
The total data rate is about 30 Mbit/s.

From this input, the encoder creates a stream of coded packets that are distributed on two simulated paths.
We model the paths as LTE and WiFi channel according to the description in Section \ref{sec:setup} and \cite{Partov:LeithTrace}.
As specific parameters, we choose 10\% loss and 20 ms latency for the LTE path and 20\% loss with 10 ms latency for the WiFi path.
The throughput of each path should be high enough to deliver the full stream including coded packets to allow a comparison with the multipath scheme.
In our simulation, we set the throughput of each path to 40 Mbit/s.
This way we can compare the delay performance of a single path to the combination of both paths.
The coded packets will be received by the decoder.
After decoding the original input packets will be retrieved and delivered to the output in the correct order.
This means, decoded packets can be further delayed if the decoder waits for one of the previous packets.
When the packet is delivered in order, the delay of the packet is measured.

\begin{figure}
    \centering
    \begin{subfigure}[b]{\columnwidth}
        \includegraphics{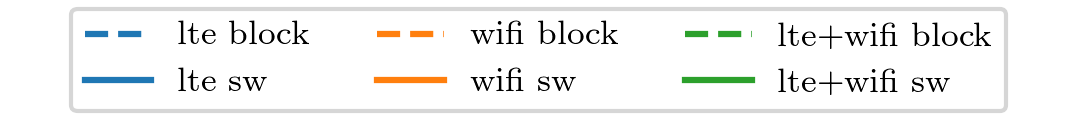}
        \includegraphics{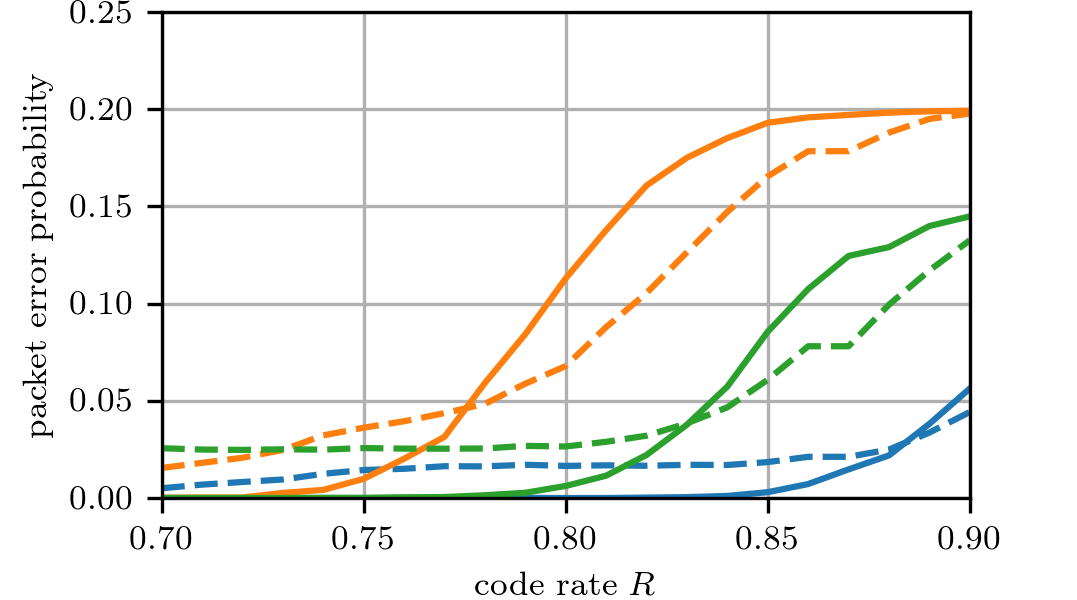}
        \caption{sliding window $w_e=64$, block code $g=64$}
    \end{subfigure}
    \begin{subfigure}[b]{\columnwidth}
        \includegraphics{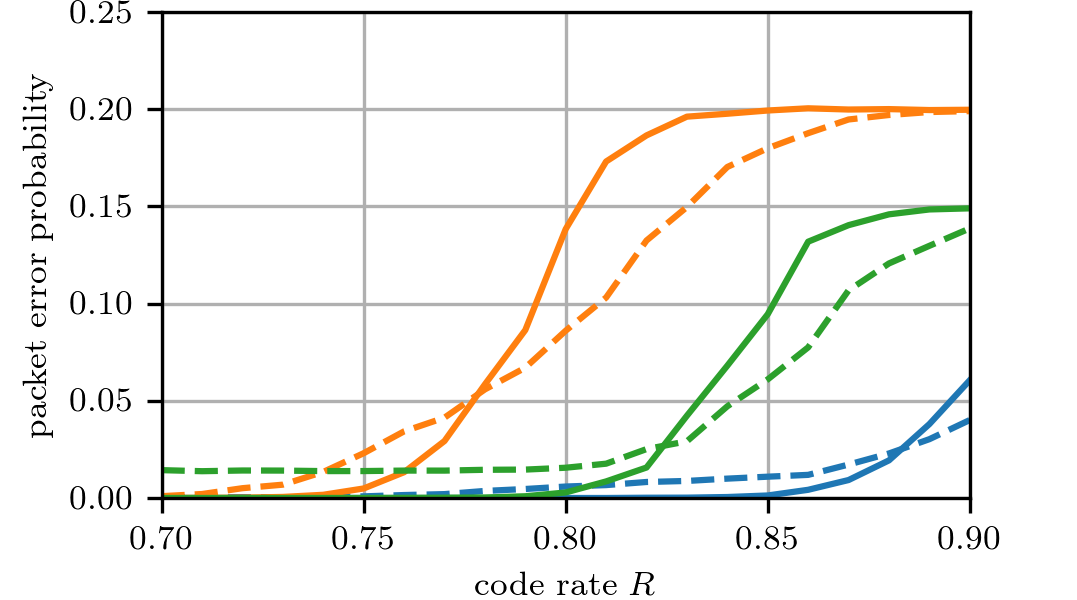}
        \caption{sliding window $w_e=96$, block code $g=96$}
    \end{subfigure}
    \begin{subfigure}[b]{\columnwidth}
        \includegraphics{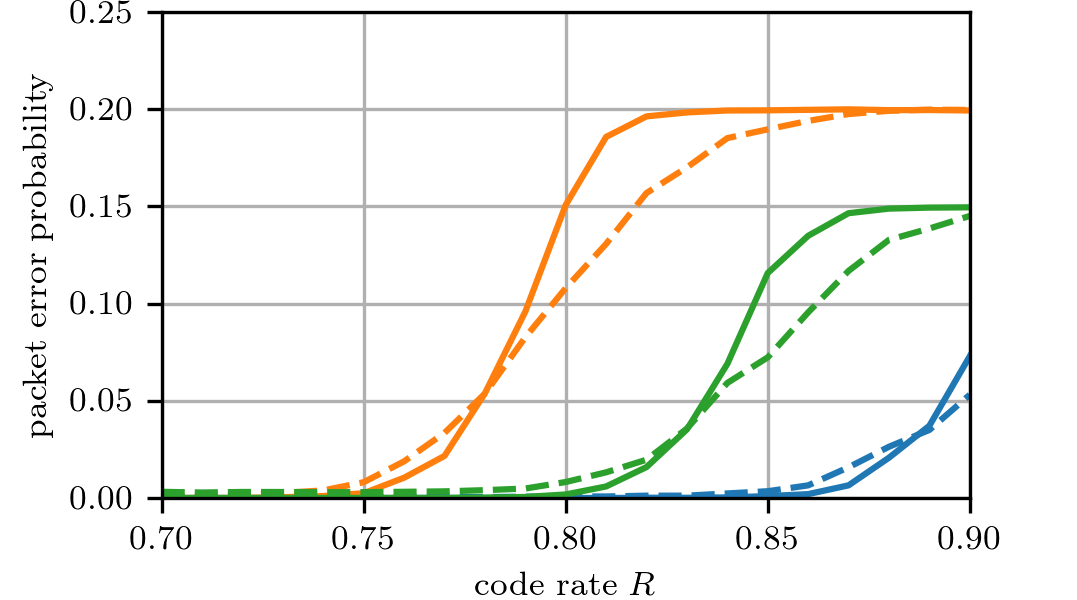}
        \caption{sliding window $w_e=128$, block code $g=128$}
    \end{subfigure}
    \caption{Packet loss probability for block codes and sliding window}
    \label{fig:loss}
\end{figure}

In the simulation, we compare a RLNC block code to the sliding window code.
The implementation of the codes is based on the work in \cite{crlnc}.
We evaluate three scenarios: LTE, WiFi, and combined LTE+WiFi.
We vary the code rate $R$ to see the relation to the residual losses and the in-order delay.
Additionally, we vary the size of the encoding window $w_e$ for the sliding window code and the generation size $g$ for the block code.
The decoding buffer size is chosen according to the analysis presented in Section \ref{sec:buffer}.
In each scenario, we simulate the transmission of 10000 packets and repeat for 20 runs.

\figurename~\ref{fig:loss} shows the packet loss probability for the scenario with different window sizes and code rates.
With decreasing code rate the packet loss probability decreases.
As expected from the configuration of the paths, LTE has the least losses.
For example with the sliding window with $w_e=96$, it reaches 99.99\% reliability at the code rate 0.83.
Single path WiFi requires a code rate of 0.72 to recover all losses.
The performance of the combined paths is between the single path performance of LTE and WiFi and reaches 99.99\% losses at the code rate 0.78.
Given the round-robin scheduling of the packets, this is the expected behavior.
Compared to the block code, the sliding window is less effective in higher code rates, but more effective in lower code rates.
Most importantly, the sliding window code reaches high reliability earlier than the block code.

The loss probability can also be decreased by increasing the encoding window size.
With an increased encoding window size or generation size, the code can recover longer bursts of errors.
Each coded packet is a combination of more source packets and can be potentially used to recover more error scenarios.
Especially the block code can significantly improve its performance with an increased generation size.

\begin{figure}
    \centering
    \begin{subfigure}[b]{\columnwidth}
        \includegraphics{coderate_legend}
        \includegraphics{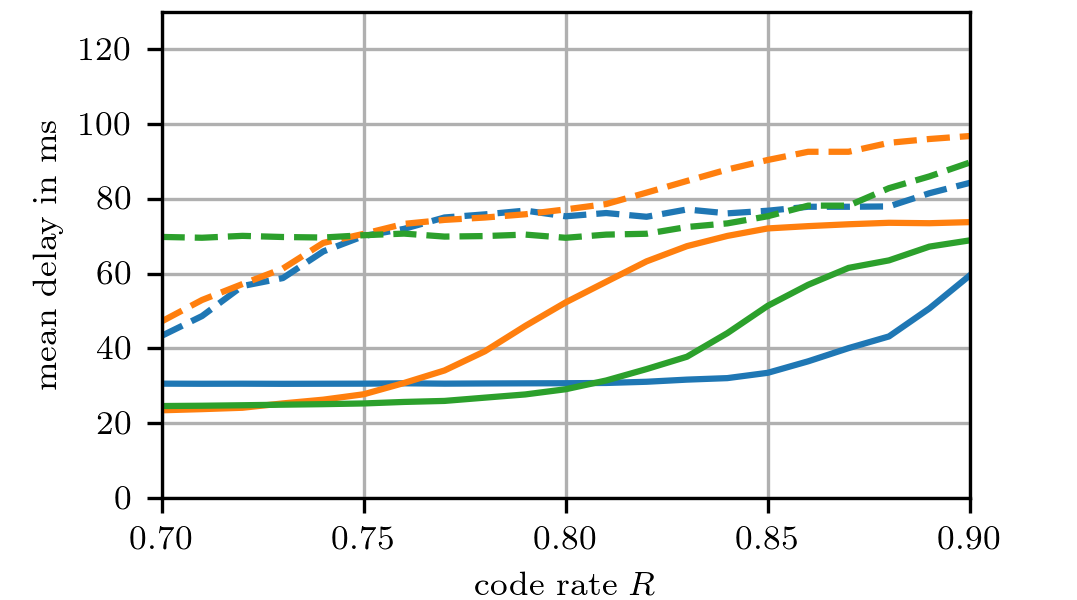}
        \caption{sliding window $w_e=64$, block code $g=64$}
    \end{subfigure}
    \begin{subfigure}[b]{\columnwidth}
        \includegraphics{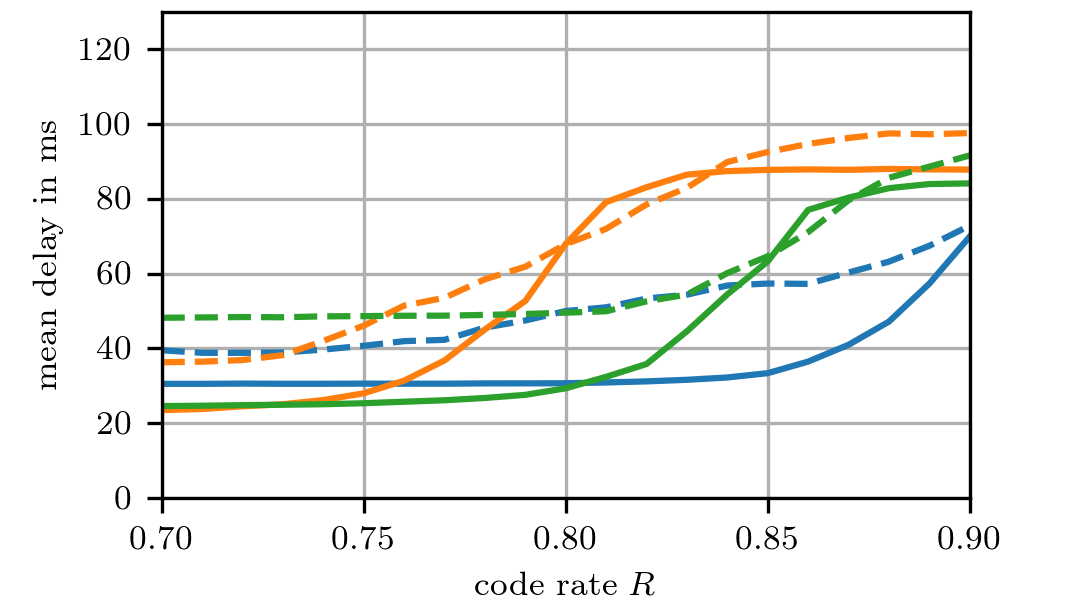}
        \caption{sliding window $w_e=96$, block code $g=96$}
    \end{subfigure}
    \begin{subfigure}[b]{\columnwidth}
        \includegraphics{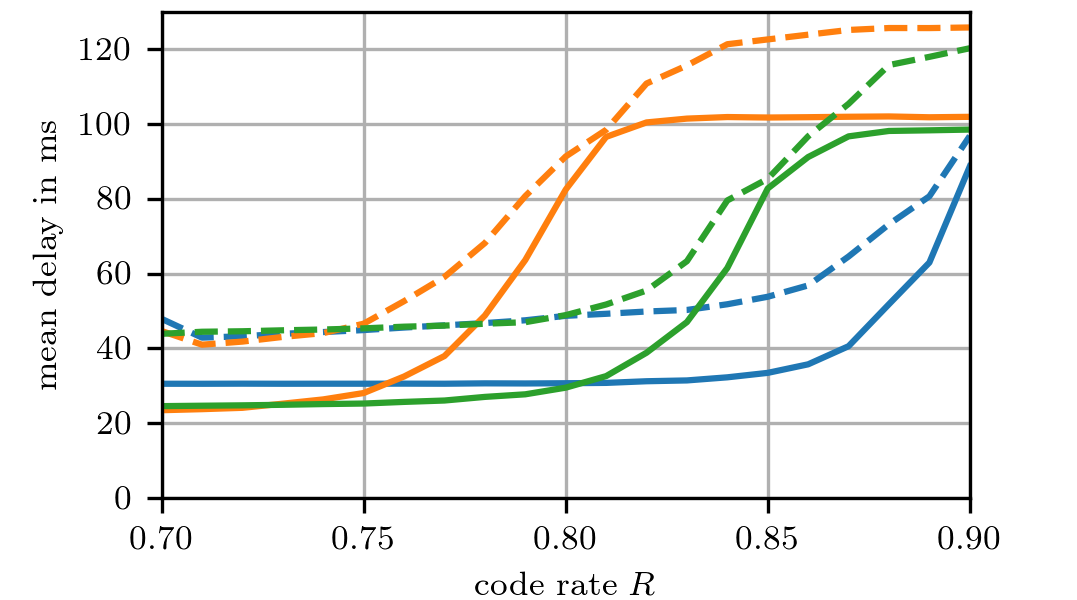}
        \caption{sliding window $w_e=128$, block code $g=128$}
    \end{subfigure}
    \caption{Mean delay for block codes and sliding window}
    \label{fig:delay}
\end{figure}

With an increased encoding window or generation size, a coded packet can recover an error that happened in an earlier packet.
That means the packet is recovered, but with a high delay.
\figurename~\ref{fig:delay} shows the mean delay for the scenarios.
In general, the delay of the sliding window is lower than for the block code.
This is consistent with the findings in other works such as \cite{crlnc,RocaSW}.
With a decreased code rate the delay improves.
The main influence on this behavior is the reduced head-of-line blocking when more packets can be recovered.
The biggest decrease in delay is up to the point where close to zero losses are achieved.
Further decreasing the code rate has only a minor influence on the delay.

The main finding is that in the code rates between 0.75 and 0.8 the combination of LTE and WiFi outperforms the single path scenarios.
In this area, the combination of LTE and WiFi operates at near-zero losses and has up to 5ms (18\%) lower delay compared to LTE.
The WiFi path requires a lower code rate to achieve comparable reliability and delay.
Table~\ref{tab:results} shows the delays at the points where the paths reach full reliability for $w_e=96$.
The reason for this behavior is the increased throughput of the combination of both paths.
By combining the paths the queuing time of the individual packets can be reduced. 
Even though WiFi has relatively high losses, it is still beneficial to use it to reinforce the LTE path.

\begin{table}[t]
    \centering
    \caption{Highest code rate and respective delay to reach 99.99\% losses with $w_e=96$.}
    \begin{tabular}{|c|c|c|}
        \hline
        Path & Code rate & Delay \\
        \hline
        LTE & 0.83 & 31.3ms \\
        WiFi & 0.72 & 24ms \\
        LTE+WiFi & 0.78 & 26.5ms \\
        \hline
    \end{tabular}
    \label{tab:results}
\end{table}

\section{Conclusion}
\label{sec:Conc}
We have performed an evaluation of the multipath sliding window network coding scheme using a simulation for two paths and showed that using both the paths is better compared to using only the best path in terms of latency. Also for the same code
rate and window size the sliding window code outperforms the block code.
We have carried out the evaluation for a range of code rates and encoding window sizes and showed that the increased window size reduces the packet losses for the multipath similar to single path case. Our scheme does not rely on feedback and uses only a fixed, low window size and shows better delay performance compared to \cite{CloudM16}. The cumulative data rate obtained from the multipath communication is converted into the gain in latency for the bursty traffic of a real-time video stream. We have also obtained bounds for the decoding window size. The proposed scheme achieves better delay performance with a higher code rate (lower redundancy) compared to the individual paths and thus reaches low delay requirements more efficiently than a single path solution.

As a future direction, we would like to look at the effect of recoding, which is a salient feature of RLNC, at the different paths. Another topic is the scalability of the scheme to more than two paths; we expect that the characteristic of the paths with lowest and highest delay will have the prominent effect on the performance due to delay asymmetry and the losses will average out over the paths.

\section*{Acknowledgements}

This article is based upon work supported by Ericsson, Deutsche Telekom and the German Federal Ministry for Education and Research (TacNet, FKZ: 16KIS0736)


\bibliographystyle{IEEEtran}
\bibliography{BibMPSW}

\end{document}